\documentclass[11pt]{article}
\usepackage[left=1in,top=1in,right=1in,bottom=1in]{geometry}
\usepackage{times}

\usepackage{graphicx}
\usepackage{url}
\usepackage{verbatim}
\usepackage{citesort}
\makeatletter
\g@addto@macro{\UrlBreaks}{\UrlOrds}
\makeatother

\begin{document}  

\title{My Reflections on the First Man vs. Machine \\ No-Limit Texas Hold 'em Competition
\thanks{The competition was organized by Professor Tuomas Sandholm, and the agent was created by Noam Brown, Sam Ganzfried, and Tuomas Sandholm. This article contains the author's personal thoughts on the event. Some of the work described in this article was performed while the author was a student at Carnegie Mellon University before the completion of his PhD. The article reflects the views of the author alone and not necessarily those of Carnegie Mellon University. The work done at Carnegie Mellon University was supported by the National Science Foundation under grants IIS-1320620, IIS-0964579, and CCF-1101668, as well as XSEDE computing resources provided by the Pittsburgh Supercomputing Center.}}

\author{Sam Ganzfried \\ sam.ganzfried@gmail.com}

\date{\vspace{-5ex}}

\maketitle

\begin{abstract} 
The first ever human vs. computer no-limit Texas hold 'em competition took place from April 24--May 8, 2015 at River's Casino in Pittsburgh, PA. In this article I present my thoughts on the competition design, agent architecture, and lessons learned.
\end{abstract}

\section{Introduction}
The first ever human vs. computer no-limit Texas hold 'em competition took place from April 24--May 8, 2015 at River's Casino in Pittsburgh, PA, organized by Carnegie Mellon University Professor Tuomas Sandholm. 20,000 hands of two-player no-limit Texas hold 'em were played between the computer program ``Claudico'' and four of the top human specialists in this variation of poker, Dong Kim, Jason Les, Bjorn Li, and Doug Polk (so 80,000 hands were played in total).\footnote{Doug Polk tweeted a list on 2/28/2015 ranking himself at number one, Kim number two, Li number three, and Les (according to speculation on his screenname) within the top ten,~\url{https://twitter.com/DougPolkPoker/status/571647246074163201}. Several other players have also created lists placing Polk at number one (e.g., Nick Frame tweeted one on 9/28/2014,~\url{https://twitter.com/TCfromUB/status/516396810433486848}). While these rankings are largely subjective, they are based on some objective factors; e.g., if player A beats player B over a significant sample of hands, or if player A is willing to play against player B but player B refuses to play against player A (i.e., by leaving the table when player A sits in against him), then these indicate an advantage of player A over player B. If one player contests the ranking and believes he is better than someone ranked higher, then a challenge can ensue (e.g., Kim and Frame played a challenge match in February 2015,~\url{https://www.pokerstars.com/en/blog/2015/dong-donger-kim-kyu-and-nick-tcfromub-frame-on-their-unique-heads-up-challenge-up-challenge-154091.shtml}).}

To evaluate the performance, we used ``duplicate'' scoring, in which the same hands were played twice with the cards reversed to reduce the role of luck (and thereby the variance).\footnote{For example, suppose human A has pocket aces and the computer has pocket kings, and A wins \$5,000. This would indicate that the human outplayed the computer. However, suppose human B has the pocket kings against the computer's pocket aces in the identical situation and the computer wins \$10,000. Then, taking both of these results into account, an improved estimator of performance would indicate that the computer outplayed the human, after the role of luck in the result was significantly reduced.} Each human was given a partner, who played the identical hands against Claudico with the cards reversed. Polk was paired with Les, and Kim was paired with Li. The players played in two different rooms of the casino simultaneously, with one player from each of the pairings in each room.
In total, the humans ended up winning the match by 732,713 chips, which corresponds to a win rate of 9.16 big blinds per 100 hands (BB/100),\footnote{The small blind (SB) and big blind (BB) correspond to initial investments, or ``antes'' of the players. In the match, the SB was 50 chips and the BB was 100 chips.} a common metric used to evaluate performance in poker. This was a relatively decisive win for the humans and was statistically significant at the 90\% confidence level, though it was not statistically significant at the 95\% level.\footnote{To put these results into some perspective, Dong Kim won the challenge described above against Nick Frame by 13.87 BB/100 (he won by \$103,992 over 15,000 hands with blinds SB=\$25, BB=\$50),~\url{http://www.pokergurublog.com/content/donger-kim-wins-heads-challenge-against-tcfromub}, and Doug Polk defeated Ben Sulsky in another high-profile challenge match by 24.67 BB/100 (he won by \$740,000 over 15,000 hands with blinds SB = \$100, BB = \$200),~\url{http://www.pokernews.com/news/2013/10/doug-polk-defeats-ben-sulsky-16618.htm}.}    

The chips were just a placeholder to keep track of the score and did not represent real money; the humans were paid at the end from a prize pool of \$100,000 which had been donated from River's Casino and Microsoft Research. The human with the smallest profit over the match received \$10,000, while the other humans received \$10,000 plus additional payoff in proportion to the profit above the lowest profit. Formally, let $x_1,x_2,x_3,x_4$ denote the profits of the four humans from highest to smallest, and let $p_i$ denote the corresponding payoffs. Then
\begin{eqnarray}
\mbox{If } x_1 &> &x_4 \\
p_1 &= &\$10,000 + \$60,000 \cdot \frac{x_1-x_4}{x_1+x_2+x_3-3x_4} \\
p_2 &= &\$10,000 + \$60,000 \cdot \frac{x_2-x_4}{x_1+x_2+x_3-3x_4} \\
p_3 &= &\$10,000 + \$60,000 \cdot \frac{x_3-x_4}{x_1+x_2+x_3-3x_4} \\
p_4 &= &\$10,000 \\
\mbox{Else } && \\
p_1 &= &p_2 = p_3 = p_4 = \$25,000
\end{eqnarray}
This scheme ensured that all players received at least \$10,000 and that payoffs were increasing in profit, giving each human a financial incentive to try their best individually.

While this was the first man vs. machine competition for the no-limit variant of Texas hold 'em, there had been two prior competitions for the limit variant. In the limit variant all bets are of a fixed size, while in no-limit bets can be of any number of chips up to the amount remaining in a player's stack (the stacks are reset to a fixed amount of 200 big blinds at the start of each hand). Thus, the game tree for no-limit has a much larger branching factor and is significantly larger; there are $10^{165}$ nodes in the game tree for no-limit, while there are around $10^{17}$ nodes for limit~\cite{Johanson13:Measuring}. In 2007 a program called Polaris that was created by researchers at the University of Alberta played four duplicate 500-hand matches against human professionals. The program won one match, tied one, and lost two, thus losing the match overall. In 2008 an improved version of Polaris competed against six human professionals in a second match, this time coming out victorious (three wins, two losses, and one tie). There have also been highly-publicized man vs. machine competitions for other games; for example, chess program Deep Blue lost to human expert Garry Kasparov in 1996 and beat him in 1997, and Jeopardy agent Watson defeated human champions in 2011.

Claudico is Latin for ``I limp.'' Limping is the name of a specific play in poker. After the initial antes have been paid, the first player to act is the small blind and he has three available actions; fold (forfeit the pot), call (match the big blind by putting in 50 chips more), or raise by putting in additional chips beyond those needed to call (a raise can be any integral amount from 200 chips up to 20,000 chips in this situation). The second option of just calling is called ``limping'' and has traditionally been viewed as a very weak play only made by bad players. In one popular book on strategy, Phil Gordon writes, ``\textbf{Limping is for Losers.} This is \emph{the most important fundamental} in poker---for every game, for every tournament, every stake: If you are the first player to voluntarily commit chips to the pot, open for a raise. Limping is inevitably a losing play. If you see a person at the table limping, you can be fairly sure he is a bad player. Bottom line: If your hand is worth playing, it is worth raising''~\cite{Gordon11:Phil}. Claudico actually limps close to 10\% of its hands, and based on discussion with the human players who did analysis it seems to have profited overall from the hands it limped. Claudico also makes several other plays that challenge conventional human poker strategy; for example it sometimes makes very small bets of 10\% of the pot, and sometimes very large all-in bets for many times the pot (e.g., betting 20,000 into a pot of 500). By contrast, human players typically utilize a small number of bet sizes, usually between half pot and pot.

\section{Agent Architecture}
\begin{figure}[!ht]
\centering
\includegraphics[scale = 0.3]{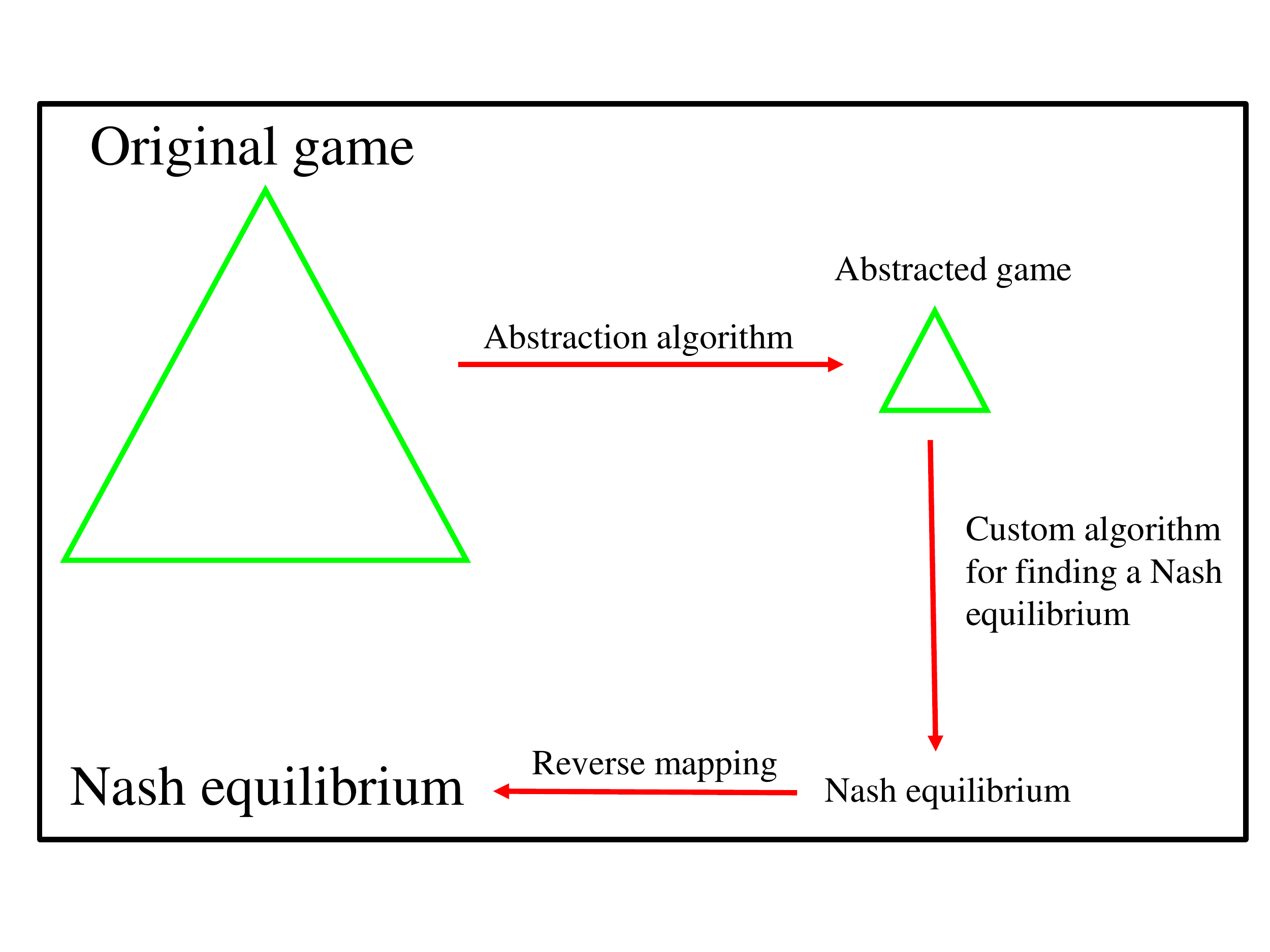}
\caption{Leading paradigm for solving large games.}
\label{fi:leading-paradigm}
\end{figure}

Claudico was an improved version of an earlier agent called Tartanian7 that came in first place in the 2014 AAAI computer poker competition, beating each opposing agent with statistical significance. The architecture of that agent has been described in detail in a recent paper~\cite{Brown15:Hierarchical}. At a very high level, the design of the agent follows the three-step procedure depicted in Figure~\ref{fi:leading-paradigm}, which is the leading paradigm used by many of the strongest agents for large games.

In the first step, the original game is approximated by a smaller \emph{abstract game} that hopefully retains much of the strategic structure of the initial game. The first abstractions for two-player Texas hold 'em were manually generated~\cite{Shi01:Abstraction,Billings03:Approximating}, while current abstractions are computed algorithmically~\cite{Gilpin06:Competitive,Gilpin07:Better,Gilpin08:Heads-up,Waugh09:Practical,Johanson13:Evaluating}. For smaller games, such as Rhode Island hold 'em, abstraction can be performed losslessly, and the abstract game is actually isomorphic to the full game~\cite{Gilpin07:Lossless}. However, for larger games, such as Texas hold 'em, we must be willing to incur some loss in the quality of the modeling approximation due to abstraction. 

The second step is to compute an $\epsilon$-equilibrium in the smaller abstracted game, using a custom iterative equilibrium-finding algorithm such as counterfactual regret minimization (CFR)~\cite{Zinkevich07:Regret} or a generalization of Nesterov's excessive gap technique~\cite{Hoda10:Smoothing}.

The final step is to construct a strategy profile in the original game from the approximate equilibrium of the abstracted game by means of a \emph{reverse mapping} procedure. When the action spaces of the original and abstracted games are identical, this step is often straightforward, since the equilibrium of the abstracted game can be played directly in the full game. However, even in this simplified setting often significant performance improvements can be obtained by applying a nontrivial reverse mapping. Several procedures have been shown to significantly improve performance that modify the action probabilities of the abstract equilibrium strategies by placing more weight on certain actions~\cite{Ganzfried12:Strategy,Brown15:Hierarchical}. These \emph{post-processing} procedures are able to achieve robustness against limitations of the abstraction and equilibrium-finding phases of the paradigm. 

When the action spaces of the original and abstracted games differ, an additional procedure is needed to interpret actions taken by the opponent that are not allowed in the abstract game model. Such a procedure is called an \emph{action translation mapping}. The typical approach for performing action translation is to map the opponent's action to a nearby action that is in the abstraction (perhaps probabilistically), and then respond as if the opponent had taken this action. 

\begin{figure}[!ht]
\centering
\includegraphics[scale = 0.3]{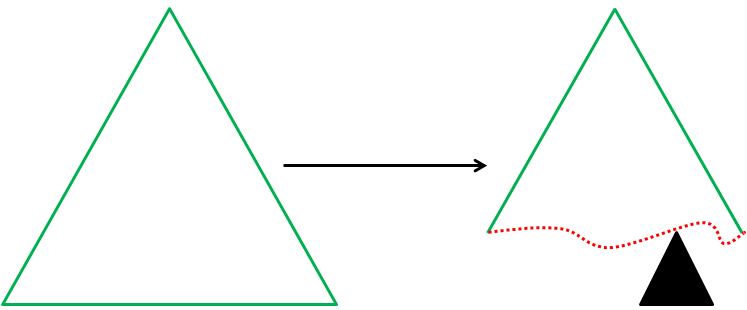}
\caption{Endgame solving (re-)solves the relevant endgame that we have actually reached in real time to a greater degree of accuracy than in the offline computation.}
\label{fi:endgame-solving}
\end{figure}

An additional crucial component of Claudico, that was not present in Tartanian7 due to a last-minute technical difficulty (thought a version of it was present in prior agent Tartanian6), is an approach for real-time computation of solutions in the part of the game tree that we have reached to a greater degree of accuracy than in the offline computation, called \emph{endgame solving}, which is depicted in Figure~\ref{fi:endgame-solving}~\cite{Ganzfried15:Endgame}.
At a high level, endgame solving works by assuming both agents follow the precomputed approximate equilibrium strategies for the \emph{trunk} portion of the game prior to the endgame; then the endgame induced by these trunk strategies is solved, using Bayes' rule to compute the input distributions of players' private information leading into the endgame. In general, such a procedure could produce a non-equilibrium strategy profile (even if the full game has a unique equilibrium and a single endgame); for example, in a sequential version of rock-paper-scissors where player 1 acts and then player 2 acts without observing the action taken by player 1, if we fix player 1 to follow his equilibrium strategy of randomizing equally among all three actions, then any strategy for player 2 is an equilibrium in the resulting endgame, because each one yields her expected payoff 0. In particular, the equilibrium solver could output the pure strategy Rock for her, which is clearly not an equilibrium of the full game. On the other hand, endgame solving is successful in other games; for example in a game where player 1 first selects an action $a_i$ and then an imperfect-information game $G_i$ is played, we could simply solve the $G_i$ corresponding to the action $a_i$ that is actually taken, provided that the $G_i$ are independent and no information sets extend between several $G_i$.
Furthermore, endgame solving has been previously demonstrated to improve performance empirically against strong computer programs in no-limit Texas hold 'em~\cite{Ganzfried15:Endgame}. 

We used the endgame solver to compute our strategies in real time for the final betting round of each hand, called the river.\footnote{There are (up to) four betting rounds in a hand of Texas hold 'em poker. First both players are dealt two private cards and there is an initial round called \emph{preflop}. Then three public cards are dealt and there is the \emph{flop}. Then there is one more additional public card on the \emph{turn}, followed by one final public card in the \emph{river} betting round.} Despite the theoretical limitation of the approach, Doug Polk related to me in personal communication after the competition ended that he thought the river strategy of Claudico using the endgame solver was the strongest part of the agent.

\subsection{Offline abstraction and equilibrium computation}
Claudico's action abstraction was manually generated and consisted of sizes ranging from 0.1 pot in certain situations to all-in (wagering all of one's remaining chips). The information abstraction was computed using a hierarchical algorithm that first clustered the three-card public flop boards into public buckets, then clustered the private information states for each postflop round (i.e., flop, turn, river) separately for each public bucket (no information abstraction was performed for the preflop round)~\cite{Brown15:Hierarchical}. This hierarchical abstraction algorithm allowed us to apply a new scalable distributed version of CFR~\cite{Brown15:Hierarchical}. We ran the equilibrium-finding algorithm for several months on Pittsburgh's Blacklight supercomputer using 961 cores (60 blades of 16 cores each, plus one core for the head blade, with each blade having 128 GB RAM). 

\subsection{Action translation}
For the action translation mapping, we used the pseudo-harmonic mapping, which maps a bet $x$ of the opponent to one of the nearest sizes in the abstraction $A, B$ according to the following formula, where $f(x)$ the probability that $x$ is mapped to $A$~\cite{Ganzfried13:Action}:
$$f(x) = \frac{(B-x)(1+A)}{(B-A)(1+x)}.$$ This mapping was derived from analytical solutions of simplified poker games and has been demonstrated to outperform prior approaches in terms of exploitability in simplified games, as well as the best prior approach in terms of empirical performance against no-limit Texas hold 'em agents. The mapping also satisfies several axioms and theoretical properties that the best prior mappings do not satisfy, for example it is Lipschitz continuous in $A$ and $B$, and therefore robust to small changes in the actions used in the action abstraction.

As an example to demonstrate the operation of the algorithm, suppose the opponent bets 100 into a pot of 500, and that the closest sizes in our abstraction are to ``check'' (i.e., bet 0) or to bet 0.25 pot: so $A = 0$ and $B = 0.25$. Plugging these in gives $f(x) = \frac{1}{6} = 0.167$. This is the probability we map his bet down to 0 and interpret it as a check. So we pick a random number in [0,1], and if it is above $\frac{1}{6}$ we interpret the bet as 0.25 pot, and otherwise as a check.

\subsection{Post-processing}
We used additional post-processing techniques to round the action probabilities that had been computed by the offline equilibrium-finding algorithm~\cite{Ganzfried12:Strategy}. We used a generalization of the prior approach that applied a different rounding threshold for each betting round (i.e., action probabilities below the threshold were rounded to zero and then all probabilities were renormalized), with a more aggressive (i.e., larger) threshold used for the later betting rounds, since the equilibrium-finding algorithm obtains worse convergence for those rounds due to having fewer samples. We did not apply any post-processing for ourselves on the river when using the endgame solver, and assumed neither agent used any post-processing in the generation of the trunk strategies used as inputs to the endgame solver.\footnote{It may seem somewhat strange that we applied post-processing for our own play, but assumed that no post-processing was applied for the trunk strategies entering the endgame, and that this may be problematic due to the mismatch between our own strategy and the model of it entering the endgame. We chose to do this because the endgame solving approach can be less robust if the input strategies have weight on only a small number of hands (as an extreme example, if all the weight was on one hand, then the endgame solver would assume that the other agent knew our exact hand, and the solution would require us to play extremely conservatively). The approach is much more robust if we include a small probability on many different hands before the post-processing was applied. We believed that the gain in robustness outweighed the limitation of the mismatch (in addition to the reasons given above, we already expect there to be a mismatch between the input trunk strategy for the opponent, which is based off our offline equilibrium computation, and his own actual strategy, and thus we would not be removing this mismatch completely even if we eliminated it for our own strategy).} 

\subsection{Endgame solving}
\label{se:endgame-solving}
The endgame solving algorithm consists of several steps~\cite{Ganzfried15:Endgame}. First, the joint hand-strength input distributions are computed by applying Bayes' rule to the precomputed trunk strategies, utilizing a recently developed technique that requires only a linear number of lookups in the large strategy table (while the  na\"{i}ve approach requires a quadratic number of lookups and is impractical). Then the equity is computed for each hand, given these distributions.\footnote{The \emph{equity} of a hand against a distribution for the opponent is the probability of winning plus one half times the probability of tying.} Then hands are bucketed separately for each player based on the computed equities for the given situation by applying an information abstraction algorithm. Finally an exact Nash equilibrium is computed in the game corresponding to this information abstraction and an action abstraction that had been precomputed for the specific pot and stack size of the current hand. All of this computation was done in real time during gameplay. To compute equilibria within the endgames, we used Gurobi's parallel linear program solver~\cite{Gurobi14:Gurobi} to solve the sequence-form optimization formulation~\cite{Koller94:Fast}.

\section{Problematic Hands}
\label{se:problematic}
Several notable hands stood out during the course of the competition that highlighted weaknesses of the agent, which have been singled out in a thread that was devoted entirely to the competition on the most popular poker forum, the Two Plus Two Poker Forum.\footnote{The thread discussing the event has 232,252 views and 1,609 posts as of September 23, 2015, \url{http://forumserver.twoplustwo.com/29/news-views-gossip-sponsored-online-poker-report/wcgrider-dong-kim-jason-les-bjorn-li-play-against-new-hu-bot-1526750/}. Here are links to some of the posts in the thread that relate to the hands described: 
hand~\ref{it:A4} \url{http://forumserver.twoplustwo.com/showpost.php?p=46888848&postcount=1275},
hand~\ref{it:KT} \url{http://forumserver.twoplustwo.com/showpost.php?p=46802181&postcount=831},
hand~\ref{it:blocker} \url{http://forumserver.twoplustwo.com/showpost.php?p=46773302&postcount=457}.
Note a minor clarification that Claudico invested closer to 75\% than 80\% of its stack in hand~\ref{it:KT}.}

\begin{enumerate}

\item 
\label{it:A4}
In one hand, we had A4s (ace and four of the same suit) and folded preflop after we had put in over half of our stack (the human opponent had 99). This is regarded as a bad play, since we would only need to win around 25\% of the time against the opponent's distribution for a call to be profitable at this point (we win about 33\% of the time against the hand he had). The problem was that our translation mapping mapped the opponent's raise down to a smaller size, which caused us to look up a strategy for ourselves that had been computed thinking that the pot size was much smaller than we thought it was (we thought we had invested around 7,000 when we had actually invested close to 10,000---recall that the starting stacks are 20,000). These translation issues can get magnified further as the hand develops if we think we have bet a percentage (e.g., $\frac{2}{3}$) of the (correct) size of the pot, while the strategies we have precomputed assumed a different size of the pot.

\item 
\label{it:KT}
In another hand we had KT and folded to an all-in bet on the turn after putting in about $\frac{3}{4}$ of our stack despite having top pair and a flush draw (there were three diamonds on the board and we had the king of diamonds; the opponent actually had A2 with the ace of diamonds, for a better flush draw but worse hand due to us having a pair already). The issue for this hand was that the human made a raise on the flop which was slightly below the smallest size we had in our abstraction in that situation, and we ended up mapping it down to just a call (it was just mapped down with around 3\% probability in that situation, and so we ended up getting pretty ``unlucky'' that we mapped it in the ``wrong'' direction). This ended up causing us to think we had committed far fewer chips to the pot at that point than we actually had. 

\end{enumerate}

The problem in these hands was not due simply to a flaw in the action translation mapping, or even to a flaw in the action abstraction (though of course improvements to those would be very beneficial as well); even if we had used a different translation mapping and/or used different action sizes in the abstraction, we would still have potentially sizable gaps between certain sizes of the abstraction due to the fact that we can only select so many to keep the abstraction sufficiently small so that it can be solved within time and memory limits. That means that, given the current paradigm, we will necessarily have to map bets to sizes somewhat far away with some probability, which will cause our perception of the pot size to be incorrect, as these hands indicate. This is called the ``off-tree problem,'' which has received very little study thus far. Some agents, such as versions of the agent from the University of Alberta, attempt to mitigate this problem by specifically taking actions aimed to get us back on the tree (e.g., making a bet that we would not ordinarily make to correct for the pot size disparity). However, this is problematic too, as it requires us to take an undesirable action. The endgame solving approach provides a solution to this problem by inputting the correct pot size to the endgame solving algorithm, even if this differs from our perception of it at that point due to the opponent having taken an action outside of the action abstraction. In general, real-time endgame solving could correct for many misperceptions in game state information that have been accumulated along the course of game play; however, this would not apply to the preflop, flop, and turn rounds, where we are not using endgame solving. Thus it is necessary to explore additional approaches to this problem; improved algorithms for real-time computation for the earlier rounds is a potentially promising direction, and perhaps new approaches can also be developed for addressing the off-tree problem independently of endgame solving.

We went over the log files for these two specific hands with Doug Polk in person after the competition had ended, and he agreed that our plays in both hands were reasonable had the pot size been what our computed strategies perceived it to be at that point. Of course, we both agree that the hands were both major mistakes if you include the misperception of the pot size. Even though these were only low probability mistakes due to the randomization outcome selected by the translation mapping, these types of mistakes can become a significant liability in aggregate, particularly when playing against humans who are aware of them and actively trying to exploit them. Doug alluded to this point as well in an interview after the competition.\footnote{~\url{http://www.highstakesdb.com/5793-exclusive-interview-with-no1-hunl-player-doug-wcgrider-polk.aspx}} Based on Doug's interview and subsequent conversations it seems that he views this as Claudico's biggest weakness, and it will be interesting to see what improvements can be found, and whether those can be exploited in turn by good countermeasures.

\begin{enumerate}

\setcounter{enumi}{2}

\item 
\label{it:blocker}
In one other problematic hand, we made a large all-in bet (of around 19,000) into a relatively small pot of around 1700. There were three of a suit (spades) on the board, and we had a very weak hand without a fourth spade (so our bet was a ``bluff,'' hoping the opponent would fold a stronger hand). The problem is not that we made a large bet per se, or even that we did it with a very weak hand; extremely large bets are correct and part of equilibrium strategy in certain situations,\footnote{As one example, Ankenman and Chen describe a game called the ``Clairvoyance Game'' where player 1 is dealt a winning/losing hand with probability $\frac{1}{2}$ each, and is allowed to bet any amount up to initial stack $n$ into a pot of 1; then player 2 can call or fold~\cite{Ankenman06:Mathematics}. (Player 1 knows whether he has a winning or losing hand, while player 2 does not know player 1's hand.) They analytically solve for the unique Nash equilibrium of the game, and it has player 1 betting all-in for $n$ with his winning hand, and betting all-in with some probability with his losing hand, and checking with some probability (the probability is selected to make player 2 indifferent between calling and folding); player 2 then calls and folds with some probability (which is selected to make player 1 indifferent between ``bluffing'' and checking with his losing hand). This solution holds regardless of the stack size $n$; so even if $n = 1,000,000$, it would be optimal for player 1 to bet all-in for 1,000,000 to win a pot of 1 (a sketch of Ankenman and Chen's argument with the computed equilibrium strategies also appears in~\cite{Ganzfried13:Action}). Thus, it is clear that at least in certain situations extremely large bets, both with strong and weak hands, are part of optimal strategies.} and in such situations they must be made with some weak hands as bluffs to balance with the very strong ``value'' hands or else our strategy would be too predictable (if we never bluffed, then the opponent would just fold everything except his hands that beat half of our value hands, and then the bets with the bottom half of our value hands would be unprofitable). Thus, making large bets as bluffs is needed in certain situations. The problem is that certain hands are much better suited for them than others. For example, suppose the board was JsTs4sKcQh, and suppose we could have 3c2c (three and two of clubs) vs. 3s2c (three of spades and two of clubs). Both hands are extremely weak (they produce the worst possible five-card hand); however, if we have the 3s, it actually has a subtle and very significant benefit: it significantly reduces the probability that the opponent holds an extremely strong hand (e.g., an ace-high or king-high flush) because several of the hands that would constitute that strength would contain that card, e.g., As3s and Ks3s. Thus, this would make a much better choice for our hand to make a large bet with, since he is less likely to have a hand strong enough to call, making the bluff bet more effective. Our endgame-solving algorithm described in Section~\ref{se:endgame-solving} takes this ``card removal'' factor into account to an extent, since the equities are computed for each hand against the distribution the opponent could hold given that hand; however, this does not fully take into account the card removal effect. For example, the 3c2c and 3s2c hands would both have the lowest possible equity (it would be slightly above zero only because of possible ties), and would be necessarily grouped into the same bucket by our endgame information abstraction algorithm (the worst bucket) despite the fact that they have very different card removal properties. 
\end{enumerate}

Doug Polk said that he thought the river strategy using the endgame solver overall was the strongest part of Claudico; however, he thought that utilizing the large betting sizes without properly accounting for card removal was actually a significant weakness, since we would be bluffing with non-optimal hands. We came to this conclusion ourselves as well during the competition, and for this reason decided to take out the large bets for ourselves from the endgame solver partway through the competition, since this issue is most problematic for those bet sizes (for smaller bet sizes, card removal is still important, but significantly less important since we are not just trying to ``block'' the opponent from having a small number of extremely strong hands, since he will be calling with many more hands). Interestingly, Dong Kim told me after the competition that they had conducted analysis and we were actually profiting on the large bet sizes during the time we used them, despite the theoretical issue described above. I think everyone agrees that massive ``overbets'' are part of full optimal strategies, and likely underutilized by even the best human players. But card removal is also particularly important for these sizes, and I think for an agent to use them successfully an improved algorithm for dealing with blockers/card removal would need to be developed, though I am still quite curious how well we would have performed if we continued with those sizes included in the agent.


\section{Conclusion}
It is one thing to evaluate a poker agent against other computer agents, who largely also play static approximations of equilibrium strategies; it is another to compete against the strongest human specialists, who will adapt and attempt to capitalize on even the smallest perceived weaknesses. This was the first time a no-limit Texas hold 'em agent has competed against human players of this caliber, and we really had no idea what to expect entering the competition, as previously all of our experiments had been against computer agents from the AAAI Annual Computer Poker Competition. We learned many valuable lessons that will be pivotal in developing improved agents going forward. We have highlighted the two most important avenues for future research. The first is to develop an improved approach for the ``off-tree'' problem where we make a mistake due to a misperception of the actual size of the pot after translating an action for the opponent that is not in our action abstraction. We have outlined promising agendas for attacking this problem, including improved action abstraction and translation algorithms, novel approaches for real-time computation that address the portion of the game prior to the final round, and entirely new approaches specifically geared at solving the off-tree problem independently of the other problems. And the second is to develop an improved approach for information abstraction that better accounts for card removal/``blockers'' (i.e., that accounts for the fact that us having certain cards in our hand modifies the probability of the opponent having certain hands). This issue is most problematic within the information abstraction algorithm for the endgame, where the card removal effect is most significant due to the distributions for us and the opponent being the most well defined (i.e., there is no more potential remaining in the hand due to uncertainty of public cards, and this relative certainty will likely cause the distributions to put positive weight on fewer hands), and it limits our ability to utilize large bet sizes, which have been demonstrated to be optimal in certain settings. Of course, it would be beneficial to develop an improved information abstraction algorithm that accomplishes this in the part of the game prior to the endgame as well.

At first glance it may appear that these issues are purely pragmatic and specific to poker. While one of the main goals is certainly to produce a poker agent that can beat the strongest humans in two-player no-limit Texas hold 'em, there are deeper theoretical questions related to each component of the agent that has been described. Endgame solving has been proven to have theoretical guarantees in certain games while it can lead to strategies with high exploitability in others (even if the full game has a single Nash equilibrium and just a single endgame is considered)~\cite{Ganzfried15:Endgame}. It would be interesting to prove theoretical bounds on its performance on interesting game classes, perhaps classes that include variants of poker. Empirically the approach appears to be very successful on poker despite its lack of theoretical guarantees. Recently an approach has been developed for game decomposition that has theoretical guarantees~\cite{Burch14:Solving}, however from personal communication with the authors I have learned that the approach performs worse empirically than our approach that does not have a worst-case guarantee.

The main abstraction algorithms that have been successful in practice are heuristic and have no theoretical guarantees. It is extremely difficult to prove meaningful theoretical guarantees when performing such a large degree of abstraction, e.g., approximating a game with $10^{165}$ states by one with $10^{14}$ states. There has been some recent work done on abstraction algorithms with theoretical guarantees, though that work does not scale to games nearly as large as no-limit Texas hold 'em. One line of work performs lossless abstraction, that guarantees that the abstract game is exactly isomorphic to the original game~\cite{Gilpin07:Lossless}. This work has been applied to compute equilibrium strategies in Rhode Island hold 'em, a medium-sized (3.1 billion nodes) variant of poker. Recent work has also presented the first lossy abstraction algorithms with bounds on the solution quality~\cite{Kroer14:Extensive-Form}. However, the algorithms are based on integer programming formulations, and only scale to a tiny poker game with a 5-card deck. It would be very interesting to bridge this gap between heuristics that work well in practice for large games with no theoretical guarantees, and the approaches with theoretical guarantees that have more modest scalability. 

Scalable algorithms for computing Nash equilibria have diverse applications, including cybersecurity (e.g., determining optimal thresholds to protect against phishing attacks), business (e.g., auctions and negotiations), national security (e.g., computing strategies for officers to protect airports), and medicine. For medicine, algorithms that were created in the course of research on poker~\cite{Johanson12:Finding} have been applied to compute robust policies for diabetes management~\cite{Chen12:Tractable}; recently it has been proposed that equilibrium-finding algorithms are applicable to the problem of treating diseases such as the HIV virus that can mutate adversarially~\cite{Sandholm15:Steering}.
 
For the pseudo-harmonic action translation mapping, in addition to showing that it outperforms the best prior approach in terms of exploitability in several games, we have also presented several axioms and theoretical properties that it satisfies; for example, it is Lipschitz continuous in $A$ and $B$, and therefore robust to small changes in the actions used in the action abstraction~\cite{Ganzfried13:Action}. Another mapping that has very high exploitability in several games also satisfies these axioms, and further investigation can lead to deeper theoretical understanding of this problem and potentially new improved approaches.

Even the post-processing approaches, which appear to be purely heuristic, have interesting theoretical open questions. For example, it has been shown that purification (i.e., selecting the highest-probability action with probability 1) leads to an improved performance in uniform random $4 \times 4$ matrix games using random $3 \times 3$ abstractions when playing against the Nash equilibrium of the full $4 \times 4$ game for the opponent~\cite{Ganzfried12:Strategy}. These results were based off simulations that were statistically significant at the 95\% confidence level, and it would be interesting to provide a formal proof. Furthermore, that paper provided a conjecture for the specific supports of the games for which the approach would improve or not change performance, which was also based on statistically-significant simulations. It would be interesting to prove this formally as well, and to generalize the results to games of arbitrary size. On a broader level, there is relatively little theoretical understanding for why the post-processing approaches---which one would expect to make the strategies more predictable---have been shown to be consistently successful. Surprisingly, the improvements in empirical performance do not necessarily come at the expense of worst-case exploitability, and a degree of thresholding has been demonstrated to actually reduce exploitability for a limit Texas hold 'em agent~\cite{Ganzfried12:Strategy}.

\bibliographystyle{plain}
\bibliography{C://Users/sam/Documents/Research/Refs/dairefs}

\end{document}